%% file: main.tex
\documentclass[twocolumn,trackchanges]{aastex62}

\usepackage[utf8]{inputenc}
\usepackage{xcolor}
\usepackage{hyperref}
\usepackage{amsmath,units}
\usepackage{xspace}
\usepackage{acronym}
\usepackage{url}

\definecolor{chmagenta}{rgb}{0.54, 0.17, 0.88}
\definecolor{placeholder}{rgb}{0.10, 0.50, 0.10}

\acrodef{GWTC-2}{second LIGO--Virgo Gravitational Wave Transient Catalog}
\acrodef{BBH}{binary black hole}
\acrodef{GW}{gravitational wave}
\acrodef{AGN}{active galactic nucleus}

\newcommand{\oneG}{{1\text{G}}\xspace{}}
\newcommand{\twoG}{{2\text{G}}\xspace{}}
\newcommand{\firstgen}{{1\text{G}+1\text{G}}\xspace{}}
\newcommand{\halfgen}{{1\text{G}+2\text{G}}\xspace{}}
\newcommand{\secondgen}{{2\text{G}+2\text{G}}\xspace{}}

\newcommand{\smallspin}{\textsc{Model TruncGauss}\xspace{}}
\newcommand{\zerospin}{\textsc{Model ZeroSubPop}\xspace{}}

\newcommand{\Beta}{\ensuremath{\mathrm{B}}}

\newcommand{\fret}{\ensuremath{F_\mathrm{ret}}}
\newcommand{\Msun}{\ensuremath{M_\odot}}

\newcommand{\CIERA}{Center for Interdisciplinary Exploration and Research in Astrophysics (CIERA), Department of Physics and Astronomy, Northwestern University, 1800 Sherman Avenue, Evanston, IL 60201, USA}
\newcommand{\Monash}{School of Physics and Astronomy, Monash University, VIC 3800, Australia}
\newcommand{\OzGrav}{OzGrav: The ARC Centre of Excellence for Gravitational-Wave Discovery, Clayton, VIC 3800, Australia}
\newcommand{\Caltech}{LIGO Laboratory, California Institute of Technology, Pasadena, CA 91125, USA}
\newcommand{\SUPA}{SUPA, School of Physics and Astronomy, University of Glasgow, Glasgow G12 8QQ, United Kingdom}
\newcommand{\Bham}{Institute for Gravitational Wave Astronomy and School of Physics and Astronomy, University of Birmingham, Birmingham, B15 2TT, United Kingdom}
\newcommand{\Melbourne}{School of Physics, University of Melbourne, Parkville, VIC, 3010, Australia}
\newcommand{\KICP}{Kavli Institute for Cosmological Physics, The University of Chicago, 5640 South Ellis Avenue, Chicago, Illinois 60637, USA}
\newcommand{\EFI}{Enrico Fermi Institute, The University of Chicago, 933 East 56th Street, Chicago, IL 60637, USA}
\newcommand{\Santiago}{Instituto Galego de F\'{i}sica de Altas Enerx\'{i}as, Universidade de Santiago de Compostela, E-15782 Spain}
\newcommand{\WashU}{Department of Physics, Washington University, One Brookings Drive, St. Louis, Missouri 63130, USA}
\newcommand{\Swin}{Centre for Astrophysics and Supercomputing, Swinburne University of Technology, Hawthorn, VIC 3122, Australia}
\newcommand{\ck}[1]{{#1}}

\input{updated_final_pe_macros}
\begin{document}

\title{Evidence for hierarchical black hole mergers in the second LIGO--Virgo gravitational-wave catalog}
\shorttitle{Evidence for hierarchical mergers}
\shortauthors{Kimball \textit{et al}.}

\author[0000-0001-9879-6884]{Chase~Kimball}
\correspondingauthor{Chase~Kimball}
\email{CharlesKimball2022@u.northwestern.edu}
\affiliation{\CIERA}

\author[0000-0003-2053-5582]{Colm~Talbot}
\affiliation{\Caltech}

\author[0000-0003-3870-7215]{Christopher~P~L~Berry}
\affiliation{\CIERA}
\affiliation{\SUPA}

\author[0000-0002-0147-0835]{Michael~Zevin}
\affiliation{\CIERA}
\affiliation{\KICP}
\affiliation{\EFI}

\author[0000-0002-4418-3895]{Eric~Thrane}
\affiliation{\Monash}
\affiliation{\OzGrav}

\author[0000-0001-9236-5469]{Vicky~Kalogera}
\affiliation{\CIERA}

\author[0000-0002-7387-6754]{Riccardo~Buscicchio}
\affiliation{\Bham}

\author[0000-0003-4957-6679]{Matthew~Carney}
\affiliation{\WashU}

\author[0000-0003-1354-7809]{Thomas~Dent}
\affiliation{\Santiago}

\author[0000-0001-5532-3622]{Hannah~Middleton}
\affiliation{\Swin}
\affiliation{\Melbourne}
\affiliation{\OzGrav}

\author[0000-0003-4507-8373]{Ethan~Payne}
\affiliation{\Monash}
\affiliation{\OzGrav}

\author[0000-0002-6508-0713]{John~Veitch}
\affiliation{\SUPA}

\author[0000-0003-3772-198X]{Daniel~Williams}
\affiliation{\SUPA}

\begin{abstract}
We study the population properties of merging binary black holes in the second LIGO--Virgo Gravitational-Wave Transient Catalog assuming they were all formed dynamically in gravitationally bound clusters.
Using a phenomenological population model, we infer the mass and spin distribution of first-generation black holes, while self-consistently accounting for hierarchical mergers. 
Considering a range of cluster masses, we see compelling evidence for hierarchical mergers in clusters with escape velocities $\gtrsim 100~\mathrm{km\,s^{-1}}$.
For our most probable cluster mass, we find that the catalog contains at least one second-generation merger with $99\%$ credibility.
We find that the hierarchical model is preferred over an alternative model with no hierarchical mergers (Bayes factor $\mathcal{B} > 1400$) and that GW190521 is favored to contain two second-generation black holes with odds $\mathcal{O}>700$, and GW190519, GW190602, GW190620, and GW190706 are mixed-generation binaries with $\mathcal{O} > 10$.
However, our results depend strongly on the cluster escape velocity, with more modest evidence for hierarchical mergers when the escape velocity is $\lesssim 100~\mathrm{km\,s^{-1}}$.
Assuming that all binary black holes are formed dynamically in globular clusters with escape velocities on the order of tens of $\mathrm{km\,s^{-1}}$, GW190519 and GW190521 are favored to include a second-generation black hole with odds $\mathcal{O}>1$. 
In this case, we find that $99\%$ of black holes from the inferred total population have masses that are less than $\NormNominalGaussAllMassULNinetyNine \Msun$, and that this constraint is robust to our choice of prior on the maximum black hole mass.
\end{abstract}

\keywords{
Gravitational wave sources  ---  Gravitational wave astronomy --- Astrophysical black holes  ---  Hierarchical models
}

\section{Introduction}\label{intro}

The \ac{GWTC-2} has significantly expanded our set of \ac{GW} observations~\citep{Abbott:2020niy}.
It contains a total of $46$ \ac{BBH} candidates, excluding GW190814~\citep{Abbott:2020khf} whose source could also be a neutron star--black hole binary, whereas the previous catalog only contained $10$ \acp{BBH}~\citep{LIGOScientific:2018mvr}. 
Multiple astrophysical formation channels have been suggested to explain the population of \acp{BBH}, and each of these have uncertainties in their underlying physics~\citep[e.g.,][]{Kruckow:2016tti,Rodriguez:2016kxx,TheLIGOScientific:2016htt,Klencki:2018zrz,Sasaki:2018dmp,Kumamoto:2020wqr,Tagawa:2019osr,Tang:2019qhn,DiCarlo:2020lfa,Zevin:2020gma}. 
\ac{GW} observations can constrain the relative contribution of formation channels and their uncertain physics, and as the catalog grows these constraints become more precise~\citep{Stevenson:2017dlk,Talbot:2017yur,Vitale:2015tea,Zevin:2017evb,Barrett:2017fcw,Fishbach:2018edt,Zevin:2020gbd}.

Among the \ac{GWTC-2} systems there are high-mass \acp{BBH} that have components with masses of $\gtrsim 45 \Msun$~\citep{Abbott:2020niy}, the most massive being the source of GW190521~\citep{Abbott:2020tfl}. 
Black holes of $\sim45$--$135\Msun$ are not typically expected to form via standard stellar evolution as the pair-instability process either limits the maximum mass of the progenitor star's core or completely disrupts the star entirely~\citep{Fryer:2000my,Heger:2001cd,Belczynski:2016jno,Spera:2017fyx,Farmer:2019jed, Stevenson:2019rcw,Farmer:2020xne,Woosley:2021xba}. 
Potential (non-mutually exclusive) astrophysical formation mechanisms for black holes in this mass gap include hierarchical mergers, where the remnant of a previous merger becomes part of a new binary~\citep{Miller:2002pg,Antonini:2016gqe,Gerosa:2017kvu,Rodriguez:2019huv,Yang:2019cbr,Anagnostou:2020umw,Banerjee:2020bwk,Fragione:2020nib,Mapelli:2020xeq,Fragione:2020aki}; stellar mergers, which may result in a larger hydrogen envelope around a core below the pair-instability threshold~\citep{Spera:2018wnw,Kremer:2020wtp,DiCarlo:2019fcq,Gonzalez:2020xah}; formation of black holes from Population III stars that are able to retain their hydrogen envelopes~\citep{Farrell:2020zju,Kinugawa:2020xws,Vink:2020nak}, formation via stellar triples in the field~\citep{Vigna-Gomez:2020fvw}; growth via accretion in an \ac{AGN} disk~\citep{McKernan:2012rf,Michaely:2020ogo,Secunda:2020mhd,Tagawa:2019osr}, or growth via rapid gas accretion in dense primordial clusters \citep{Roupas:2019dgx}.

Hierarchical mergers in globular clusters were considered as an origin for GW190521~\citep{Abbott:2020mjq} with inconclusive results.
However, hints of eccentricity in follow-up analyses of GW190521 add weight to this explanation~\citep{Gayathri:2020coq,Romero-Shaw:2020thy}.
To confidently identify hierarchical mergers, it is important to study events in the context of a population model that fits the mass distribution (and any mass cut-offs) for the first-generation (\oneG{}) black holes not formed through mergers~\citep{Doctor:2019ruh,Sedda:2020vwo,Tiwari:2020otp,Kimball:2019mfs}. 

We apply the population inference framework from \citet{Kimball:2020opk} to analyze the \acp{BBH} from \ac{GWTC-2}. 
This framework assumes a phenomenological population model based on simulations of metal-poor globular clusters from \citet{Rodriguez:2019huv}. 
Considering a fiducial set of globular cluster masses, we simultaneously infer the properties of the \firstgen{} \ac{BBH} population---whose remnants are second-generation(\twoG{}) black holes---and the relative merger rates of hierarchical mergers. 
The expanded catalog enables the population parameters, including the mass distribution, to be more precisely determined~\citep{Abbott:2020gyp}.
We find that several of the \acp{BBH} are likely to be the results of hierarchical mergers: the leading candidates are GW190519\_153544 (GW190519) and GW190521. 
 
In Sec.~\ref{sec:methods} we review the key components of our population inference framework; the results of this are given in Sec.~\ref{sec:results}, with additional description of the population hyperparameters in Appendix~\ref{sec:hyperparameters}, and we discuss our findings in Sec.~\ref{sec:conclusion}. 

\section{Methods}\label{sec:methods}

We perform Bayesian hierarchical inference to infer the the population properties of \acp{BBH} following \citet{Kimball:2020opk}.
We employ phenomenological models for the mass and spin distributions of \firstgen{}, \halfgen{}, and \secondgen{} BBHs merging in a dense stellar environment; see Fig.~\ref{fig:massppds} and Fig.~\ref{fig:spinppds}.
The \firstgen{} model is nearly identical to population models used in \citet{Abbott:2020gyp}: it is equivalent to the \textsc{Power Law + Peak} mass model (but omits the low-mass smoothing and adopts a Gaussian prior on the maximum mass cut-off) and is similar to the \textsc{Default} spin model. We consider two separate modifications to that spin model: one that adds a parameter that allows for a subpopulation of zero-spin \acp{BBH}, and one consisting of a truncated Gaussian with a broad prior on the mean that allows for distributions with sharp peaks at $0$ or $1$; we refer to these as \zerospin{} and \smallspin{}, respectively. 
The particulars of the mass and spin models are discussed further in Appendix~\ref{sec:hyperparameters}.

The \twoG{} black holes are assumed to be roughly twice the mass of \oneG{} black holes; the mass ratio distribution for \halfgen{} binaries is peaked around $q \sim 1/2$ while the \secondgen{} distribution is similar to the \firstgen{} model but with an increased preference for near equal-mass binaries to account for the more massive components in a strong encounter forming bound binaries~\citep{1993ApJ...415..631S,1996ApJ...467..359H,Downing:2010hq}.
The \halfgen{} and \secondgen{} spin models presume that \twoG{} black holes have dimensionless spin $\chi\approx0.67$ inherited from the orbital angular momentum of the progenitor binary~\citep{Pretorius:2005gq,Gonzalez:2006md,Buonanno:2007sv}.
The population models are described as conditional priors $\pi(\theta|\Lambda)$ where $\theta$ are the parameters of a single binary (e.g., mass and spin) while $\Lambda$ refers to the population hyperparameters describing the shape of the mass and spin distributions (e.g., the power-law index of the primary black hole mass spectrum).
Our goal is two-fold: estimate the population hyperparameters $\Lambda$ and carry out model selection to evaluate the Bayesian odds that events in \ac{GWTC-2} are formed hierarchically.

The relative rates of \firstgen, \halfgen, and \secondgen{} mergers depend upon the properties of their cluster environment as well as the masses and spins of the \ac{BBH} population. 
\ac{GW} recoil kicks may lead to remnants being ejected from a cluster; kick magnitudes are strongly dependent on progenitor spins with larger spins leading to larger kicks \citep{Campanelli:2007ew, Gonzalez:2006md,Brugmann:2007zj,Lousto:2011kp,Varma:2018aht}, as well as the mass ratio of the merging binary. 
We calculate the fraction of retained merger remnants $\fret$ given the population properties of the \oneG{} black holes and assuming a cluster described by a Plummer potential \citep{Plummer:1911zza} mass of $M_\mathrm{c}$ with Plummer radius $r_\mathrm{c}$. 
For our default cluster, we assume that $M_\mathrm{c} = 5\times10^5 \Msun$ and $r_\mathrm{c} = 1~\mathrm{pc}$, corresponding to a central escape velocity of $\sim 65~\mathrm{km\,s^{-1}}$
We assume that the relative merger rates scale as $R_\halfgen{}/R_\firstgen{} \propto \fret$,  $R_\secondgen{}/R_\firstgen{} \propto \fret^2$, with the constant of proportionality calibrated against globular cluster simulations \citep{Rodriguez:2019huv}.

For our analysis, we consider the $44$ \ac{BBH} (excluding GW190814, for which the nature of the secondary component is unknown) used in the \ac{GWTC-2} population analysis \citep{Abbott:2020gyp}. 
For GWTC-1 events, we use the same single-event posterior samples as \citet{Kimball:2020opk}, for GW190412 we use samples from \citet{Zevin:2020gxf}, for GW190521 we use the preferred samples from \citet{Abbott:2020mjq}, and for the other \ac{GWTC-2} events we use the public samples from \citet{Abbott:2020niy}.\footnote{Posterior samples for GW190412, GW190521 and the other \ac{GWTC-2} systems are available from \href{https://doi.org/10.5281/zenodo.3900546}{doi.org/10.5281/zenodo.3900546}, \href{https://doi.org/10.7935/1502-wj52}{doi.org/10.7935/1502-wj52} and \href{https://doi.org/10.7935/99gf-ax93}{doi.org/10.7935/99gf-ax93}, respectively.}
For the new \ac{GWTC-2} events we use results calculated with the \textsc{IMRPhenomD} and \textsc{IMRPhenomPv2} waveforms~\citep{Khan:2015jqa,Hannam:2013oca}.
We generate posterior samples for population hyperparameters $\Lambda$ using the nested sampler \textsc{dynesty}~\citep{Speagle:2019ivv} using the \textsc{GWPopulation} framework~\citep{Talbot:2019okv}, which takes advantage of \textsc{Bilby}~\citep{Ashton:2018jfp,Romero-Shaw:2020owr}.

\section{Application to GWTC-2}\label{sec:results}

\subsection{Inferred populations}

Applying our analysis to the $44$ \ac{BBH} candidates in \ac{GWTC-2} analyzed in \citet{Abbott:2020gyp}, we infer population hyperparameters for our mass and spin models (Fig.~\ref{fig:gauss_mass}, Fig.~\ref{fig:gauss_spin_G}, and Fig.~\ref{fig:gauss_spin_delta} in Appendix~\ref{sec:hyperparameters}). For both \smallspin{} and \zerospin{}, we find that including the \halfgen{} and \secondgen{} population components is preferred, finding Bayes factors of $5$ and $7$, respectively, in favor of including versus excluding the hierarchical components.

\begin{figure}
\includegraphics[width=0.45\textwidth]{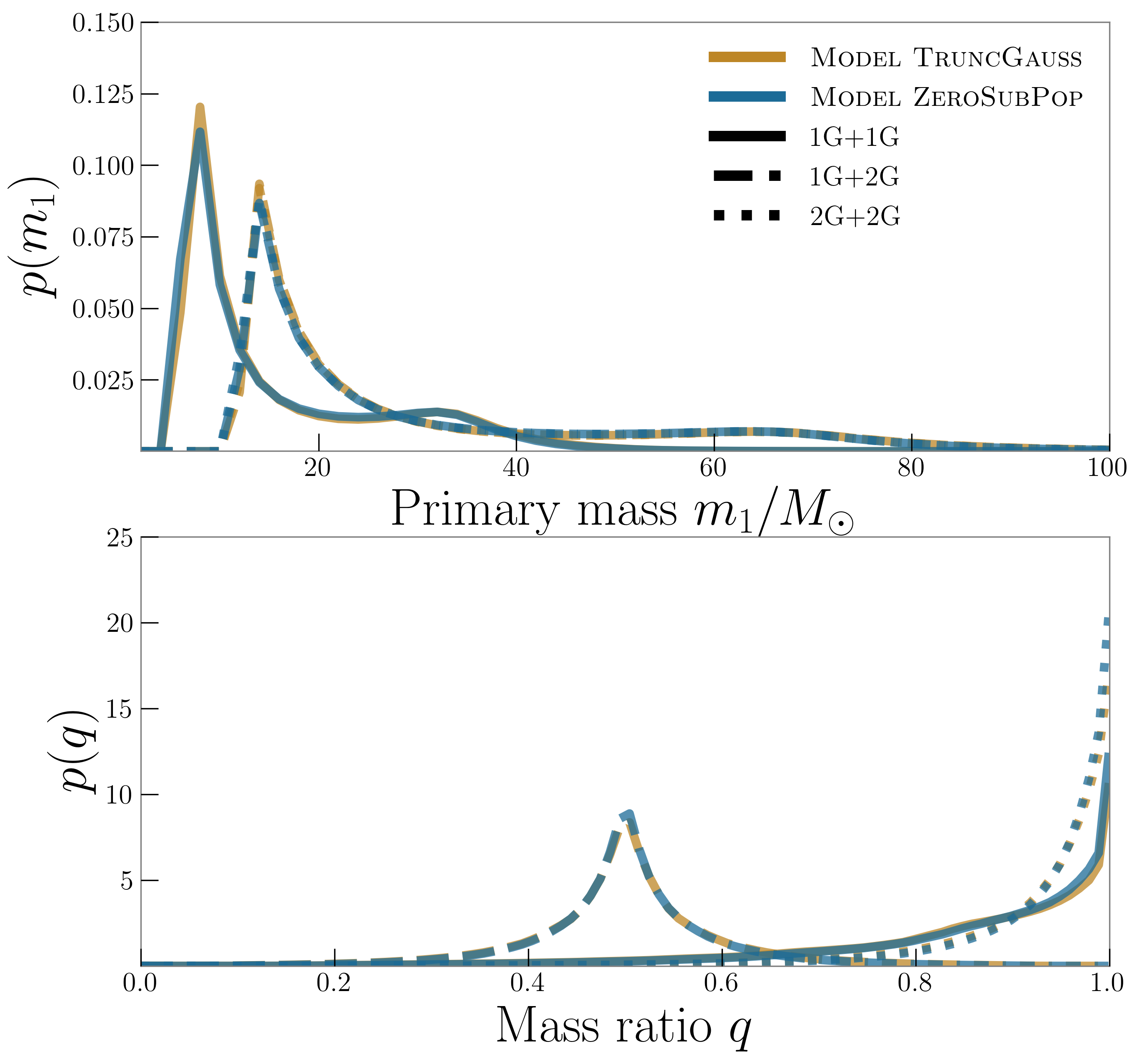}
\caption{Posterior predictive distributions for the primary mass $m_1$ and mass ratio $q$. The solid, dashed, and dotted lines indicate the \firstgen{}, \halfgen{}, and \secondgen{} distribution, respectively. In blue, we plot the distributions inferred when modeling the \oneG{} spin distribution as a non-singular Beta distribution together with a delta function at zero. In orange, we plot the distributions inferred when using a truncated Gaussian.}
\label{fig:massppds}
\end{figure}

In our inferred \oneG{} mass distribution, the mean of the Gaussian mass component is well constrained to $\mu_m=\NormNominalGaussmppMed^{+\NormNominalGaussmppPlus}_{-\NormNominalGaussmppMinus}\Msun$ and $\mu_m=\ZeroNominalGaussmppMed^{+\ZeroNominalGaussmppPlus}_{-\ZeroNominalGaussmppMinus}\Msun$ for \smallspin{} and \zerospin{}, respectively. 
For both models we recover our prior on the maximum mass cut-off $m_\mathrm{max}$. 
In Fig.~\ref{fig:massppds} and Fig.~\ref{fig:spinppds}, we plot the posterior predictive distributions for the \firstgen{}, \halfgen{}, and \secondgen{} populations. 
Using \smallspin{} (\zerospin{}), we find that $99\%$ of \firstgen{} black holes are less than $\NormNominalGaussOneGMassULNinetyNine\Msun$ ($\ZeroNominalGaussOneGMassULNinetyNine\Msun$), and $99\%$ of black holes in the total population are less than $\NormNominalGaussAllMassULNinetyNine\Msun$  ($\ZeroNominalGaussAllMassULNinetyNine\Msun$), consistent with the results of \citet{Kimball:2020opk}. These upper limits are lower than those found for the \textsc{Power Law + Peak} model in \citet{Abbott:2020gyp}, but that model does not include a high-mass hierarchical component, and requires a flatter power law to fit the heavier black holes in \ac{GWTC-2}.
Relaxing the prior on the maximum mass cut-off to a uniform prior out to $100\Msun$ (Fig.~\ref{fig:flat_200_mass}, Fig.~\ref{fig:flat_200_spin_G} and Fig. ~\ref{fig:flat_200_spin_delta}), we do not obtain stringent constraints on $m_\mathrm{max}$, but find that it peaks around $\sim 80\Msun$. In this case, we find that $99\%$ of \firstgen{} black holes are less than $\NormNominalFlatOneGMassULNinetyNine\Msun$ ($\ZeroNominalFlatOneGMassULNinetyNine\Msun$), and $99\%$ of black holes in the total population are less than $\NormNominalFlatAllMassULNinetyNine\Msun$ ($\ZeroNominalFlatAllMassULNinetyNine\Msun$).

Using \smallspin{} and \zerospin{}, $90\%$ of \firstgen{} black holes have spins less than \NormNominalGaussOneGSpinULNinety{} and \ZeroNominalGaussOneGSpinULNinety{}, respectively. 
With \zerospin{}, the fraction $\lambda_0$ of \acp{BBH} originating from the zero spin channel is constrained to be less than \ZeroNominalGaussdeltachiULNinetyNine{} at the $99\%$ credible level.

\subsection{Relative merger rates}

\begin{figure}
\includegraphics[width=0.45\textwidth]{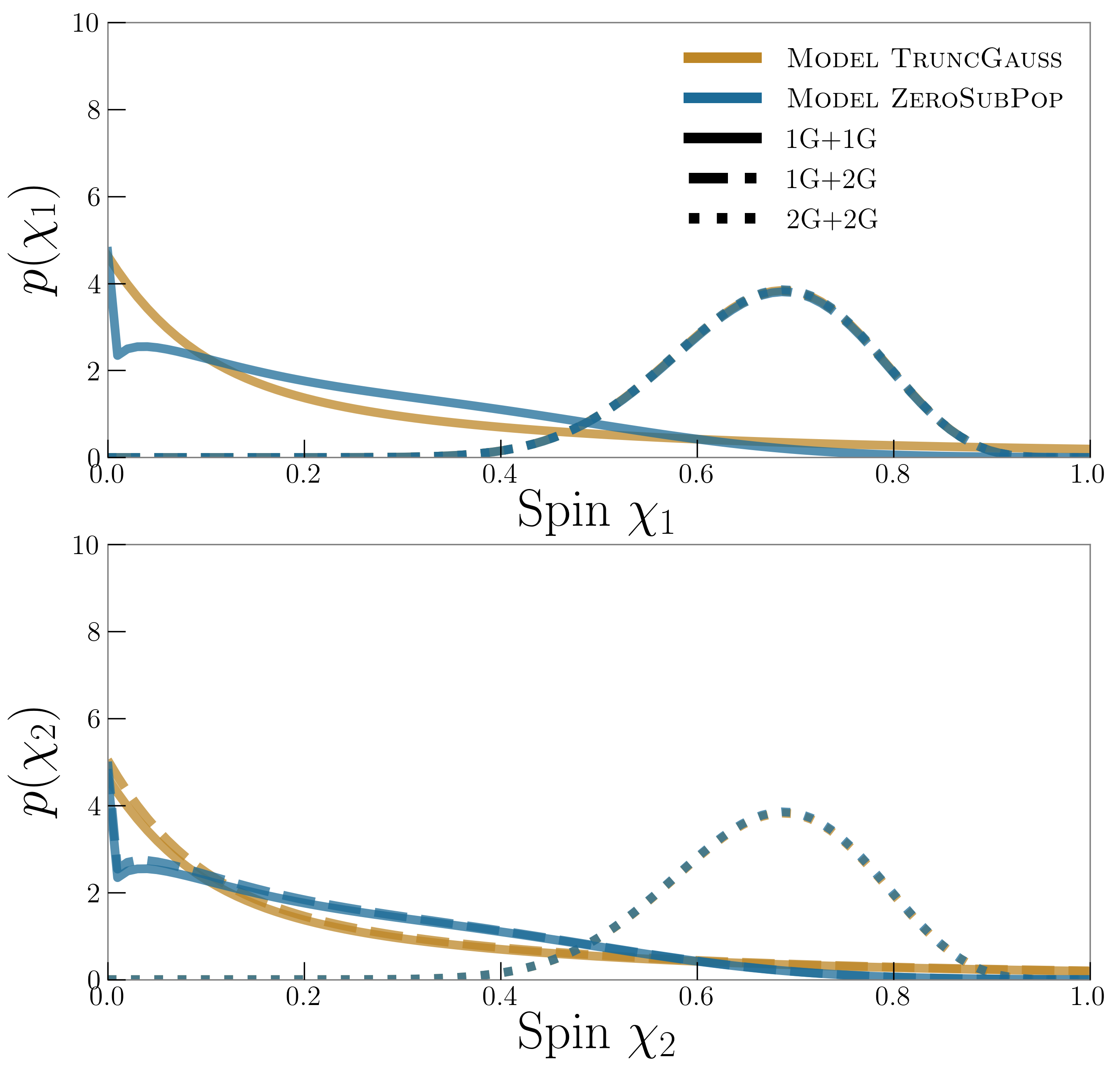}
\caption{Posterior predictive distributions for the component black hole spins. The solid, dashed and dotted lines indicate the \firstgen{}, \halfgen{}, and \secondgen{} distribution, respectively. In blue, we plot the distributions inferred when modeling the \oneG{} spin distribution as a non-singular Beta distribution together with a delta function at zero. In orange, we plot the distributions inferred when using a truncated Gaussian.}
\label{fig:spinppds}
\end{figure}

\begin{figure}
\includegraphics[width=0.45\textwidth]{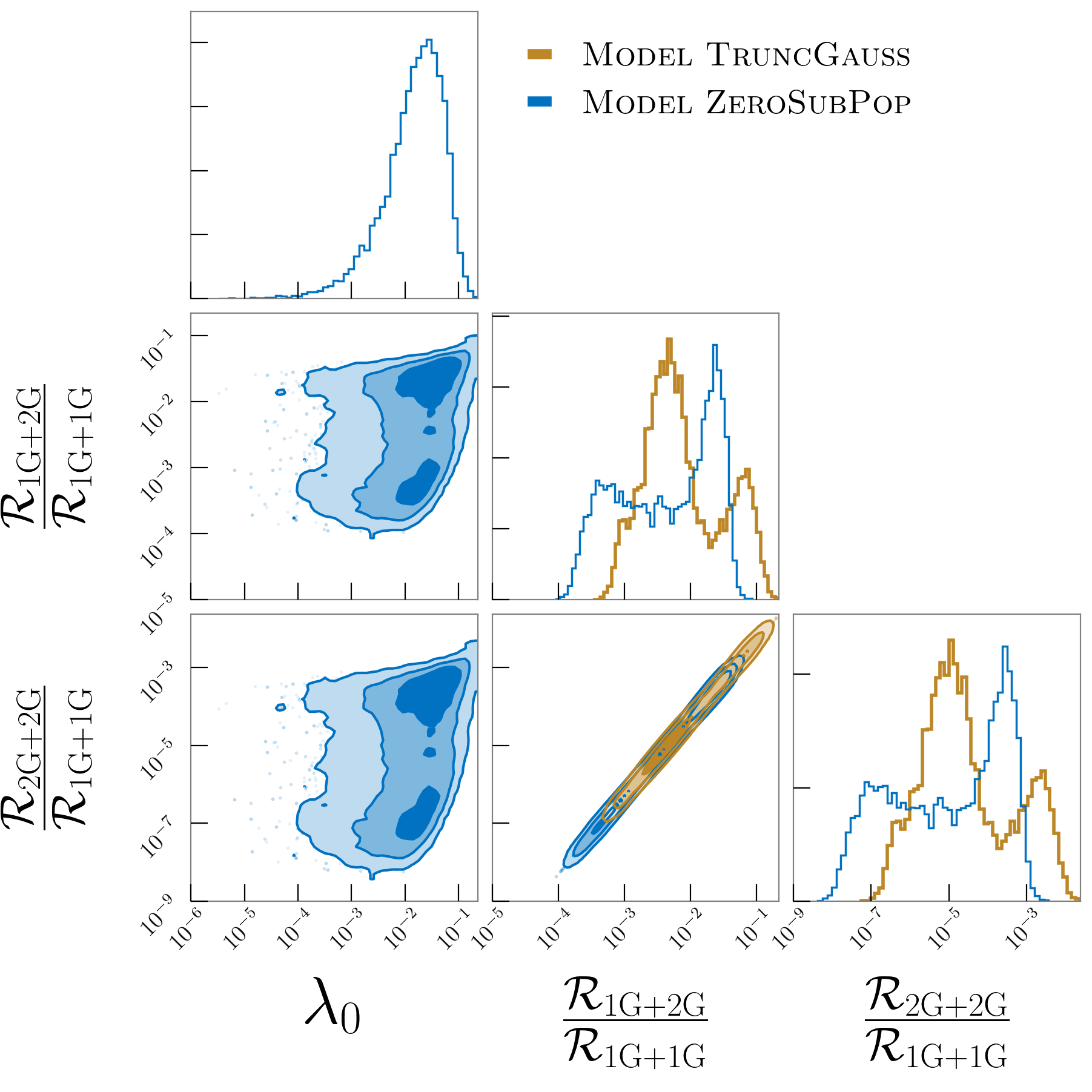}
\caption{Posteriors of the inferred branching ratios, and the fraction $\lambda_0$ of \firstgen{} black holes with zero spin for \zerospin{}. The branching ratios give the relative \halfgen{} versus \firstgen{} and \secondgen{} versus \firstgen{} merger rates. We plot the results using \smallspin{} and \zerospin{} in orange and blue, respectively.}
\label{fig:RelativeRates}
\end{figure}

\ac{GW} recoil kick velocities generally increase with the spin magnitudes of merging black holes.
Figure~\ref{fig:spinppds} illustrates that while the spin distribution inferred using \smallspin{} does not explicitly include a subpopulation at $\chi = 0$, a larger portion of the population has spins less than $\sim0.1$ than for \zerospin{}, which results in lower typical recoil velocities and hence higher relative hierarchical merger rates. In Fig.~\ref{fig:RelativeRates}, we plot the posteriors for these rates, as well as for the fraction $\lambda_0$ of \firstgen{} black holes in \zerospin{} with zero spin.  
Using \smallspin{}, we infer median relative rates $\mathcal{R}_{\halfgen}/\mathcal{R}_{\firstgen}=\NormNominalGaussBranchingHalfMed{}$ and $\mathcal{R}_{\secondgen}/\mathcal{R}_{\firstgen}=\NormNominalGaussBranchingTwoMed{}$ with $99\%$ upper limits of \NormNominalGaussBranchingHalfULNinetyNine{} and \NormNominalGaussBranchingTwoULNinetyNine{}, respectively. Taking into account selection effects, the detected population would have median relative rates of $\mathcal{R}^{\mathrm{det}}_{\halfgen}/\mathcal{R}^{\mathrm{det}}_{\firstgen}=\NormNominalGaussBranchingHalfDetMed{}$ and $\mathcal{R}_{\secondgen}/\mathcal{R}_{\firstgen}=\NormNominalGaussBranchingTwoDetMed{}$ with $99\%$ upper limits of \NormNominalGaussBranchingHalfDetULNinetyNine{} and \NormNominalGaussBranchingTwoDetULNinetyNine{}, respectively.
Using \zerospin{}, these rates become \ZeroNominalGaussBranchingHalfMed{} (\ZeroNominalGaussBranchingHalfDetMed{}) and \ZeroNominalGaussBranchingTwoMed{} (\ZeroNominalGaussBranchingTwoDetMed{}), with $99\%$ upper limits of \ZeroNominalGaussBranchingHalfULNinetyNine{} (\ZeroNominalGaussBranchingHalfDetULNinetyNine{}) and \ZeroNominalGaussBranchingTwoULNinetyNine{} (\ZeroNominalGaussBranchingTwoDetULNinetyNine{}), respectively, for the astrophysical (detected) population.
The median inferred relative rates are roughly twice those found using the same model in \citet{Kimball:2020opk}, though consistent with the upper limits reported there. The results for both models are consistent with the results of Monte Carlo modeling of black hole populations in globular clusters: 
\citet{Rodriguez:2019huv} found that the $\approx 14 \%$ of merging \acp{BBH} from the underling population in their models contain \twoG{} black holes in the extreme case where all \oneG{} black holes have zero spin (this fraction drops to $\lesssim 1\%$ when they increase \oneG{} black hole spins to $\chi=0.5$).

\subsection{Posterior odds for hierarchical origin}

\begin{figure*}
\includegraphics[width=\textwidth]{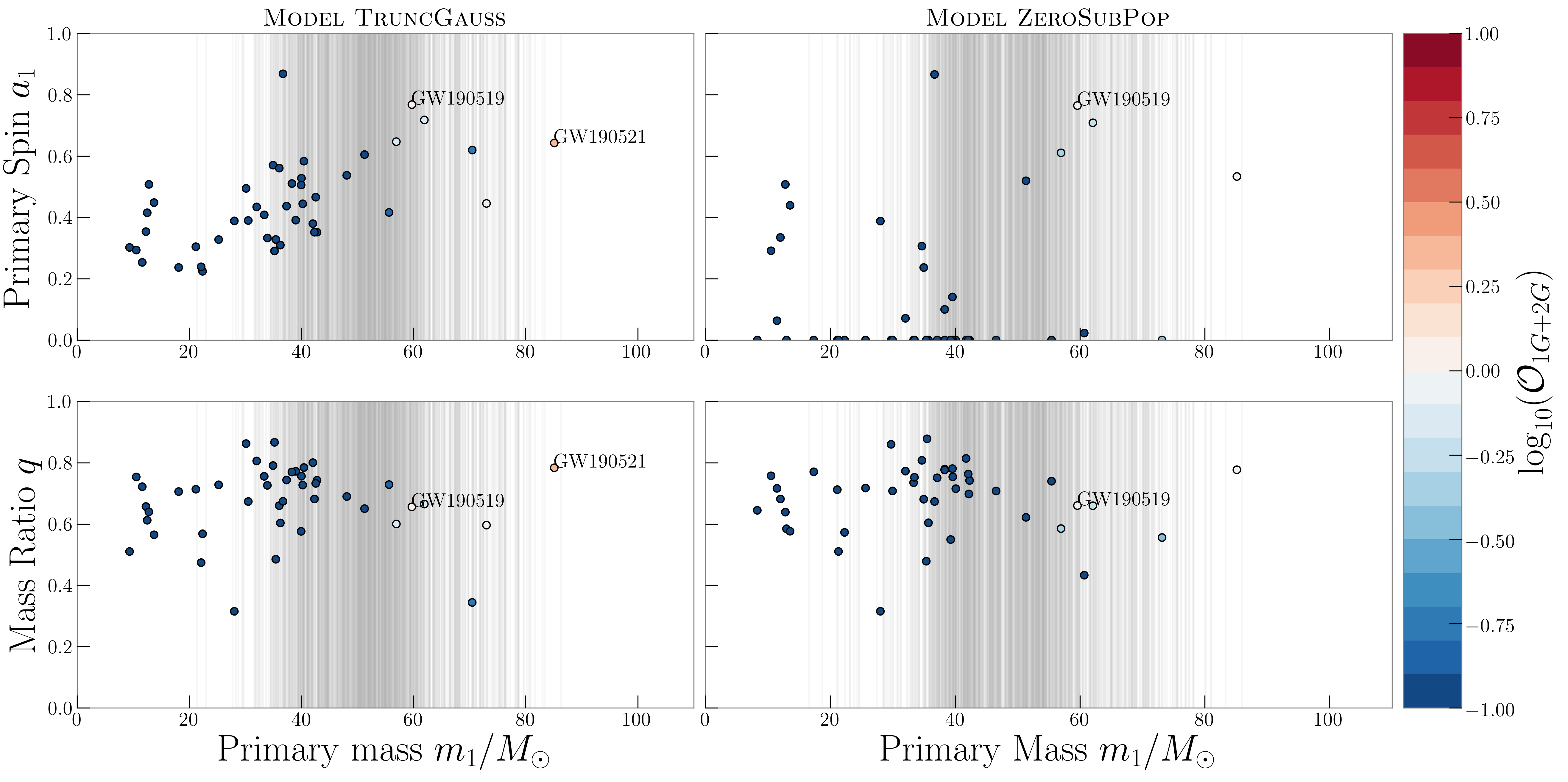}
\caption{
Odds of events in \ac{GWTC-2} having \halfgen{} versus \firstgen{} origin, as a function of the inferred median primary black hole mass, mass ratio, and primary black hole spin. The results using \smallspin{} and \zerospin{} are plotted on the left and right, respectively. The gray vertical lines are draws from the corresponding inferred posterior over the maximum \oneG{} black hole mass. 
}
\label{fig:OddsHalfG}
\end{figure*}

For each event in \ac{GWTC-2}, we calculate the posterior odds $\mathcal{O}$ in favor of hierarchical versus \firstgen{} origin. 
We plot the odds in favor of \secondgen{} versus \firstgen{} origin assuming \smallspin{} and \zerospin{} in Fig.~\ref{fig:OddsHalfG}.

Assuming \smallspin{}, we find that across all $44$ \acp{BBH} in \ac{GWTC-2} the probability that at least one binary contains a \twoG{} black hole is \NormNominalGaussPAtLeastOneTwoG $\%$. 
GW190519 and GW190521 are most likely of \halfgen{} origin, favored over a \firstgen{} origin with \NormNominalGaussGWNineteenORHalf:1 and 
2:1 odds respectively. We also favor a \secondgen{} origin over \firstgen{} for GW190521 with odds of \NormNominalGaussGWTwentyOneORTwo:1.
We find roughly even odds for GW190602\_175927 (GW190602) and GW190706\_222641 (GW190706) being of \halfgen{} origin. 
As in \citet{Kimball:2020opk}, we find that GW170729 is most likely of \firstgen{} origin, though at slightly higher odds of \ck{1:10} of being of \halfgen{} origin rather than \firstgen{}. 


Using \zerospin{}, which finds lower relative hierarchical merger rates, odds decrease across all events. 
We find that the probability that at least one binary in \ac{GWTC-2} contains a \twoG{} black hole is \ZeroNominalGaussPAtLeastOneTwoG$\%$. 
GW190519 is marginally favored to have a \halfgen{} origin with \ZeroNominalGaussGWNineteenORHalf:1 odds over a \firstgen{} origin. 
Meanwhile, GW190521 has roughly even odds of being \halfgen{} versus \firstgen{} at \ZeroNominalGaussGWTwentyOneORHalf:1 odds, and a \secondgen{} origin is disfavored to a \firstgen{} origin at \ck{1:4} odds.


\subsection{Varying cluster parameters}

Our default Plummer model is chosen as representative of a typical globular cluster environment such as those in the vicinity of the Milky Way today, where central escape velocities are on the order of tens of kilometers per second \citep{2018MNRAS.478.1520B}. 
However, globular clusters in the Milky Way may have been up to a few times more massive at formation than at present \citep{2015MNRAS.453.3278W}. 
Furthermore, hierarchical mergers may occur in a wide range of dynamical environments with significantly different escape velocities, including \ac{AGN} disks and nuclear star clusters. 
Although our phenomenological models are tuned to the results of simulations of typical present-day globular clusters \citep{Rodriguez:2019huv}, we can get an illustrative idea of how results scale with the mass and compactness of the assumed dynamical environment by varying the parameters of our simple Plummer model. 
We do not expect all \acp{BBH} to come from a single type of cluster, but our results let us explore a range of different average cluster sizes.

\begin{figure*}
\includegraphics[width=\textwidth]{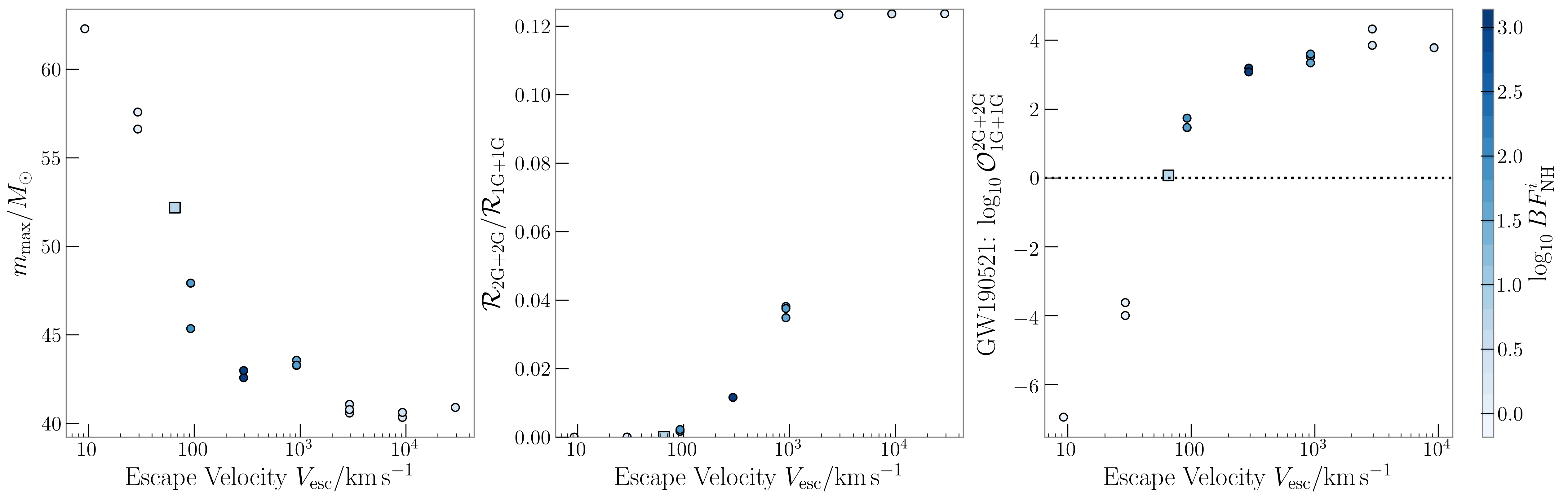}
\caption{Inferred population properties as a function of central escape velocity using \smallspin{} when considering models with Plummer masses $10^4$--$10^9 \Msun$ and radii $0.01$--$1~\mathrm{pc}$. 
In the left and middle panels, we plot the median inferred maximum black hole mass and relative \secondgen{} versus \firstgen{} merger rate. 
In the right panel, we plot the odds in favor of GW190521 being a \secondgen{} versus \firstgen{} merger. 
The points are shaded according to the Bayes factors in favor of the hierarchical model versus a model excluding hierarchical channels $\mathrm{BF}_\mathrm{NH}$. The square marker indicates our default cluster model.
}
\label{fig:vesc}
\end{figure*}

In Fig.~\ref{fig:vesc}, we show results when considering models with Plummer masses $10^4$--$10^9 \Msun$ and radii $0.01$--$1~\mathrm{pc}$; both these parameters vary cluster escape velocities and thus the retention rate of hierarchical mergers. 
At low escape velocities ($\sim 10$--$50~\mathrm{km\,s^{-1}}$), almost no \firstgen{} merger products are retained and the relative \halfgen{} and \secondgen{} rates are negligible. 
In this case, the inferred posterior on the maximum black hole mass $m_{\mathrm{max}}$ shifts away from the astrophysical prior toward higher masses in order to accommodate massive \ac{GWTC-2} events as \firstgen{} \acp{BBH}. 
As we move toward models with higher central escape velocities, the fraction of retained \firstgen{} merger products and the relative \halfgen{} and \secondgen{} rates rapidly increase, and the odds in favor of \ac{GWTC-2} events being of hierarchical origin grow. 
It is expected that the rate of hierarchical mergers strongly depends on the escape velocity of their dynamical environments \citep{HolleyBockelmann:2007eh,Moody:2008ht,Antonini:2018auk,Gerosa:2019zmo,Rodriguez:2020viw,Sedda:2020vwo,Fragione:2020han,Mapelli:2021syv}. 
We find that even a modest increase in the central escape velocity to $\sim 90~\mathrm{km\,s^{-1}}$ leads to the conclusion that GW190521 is favored to be a \secondgen{} versus \firstgen{} merger at \ck{$>10$:$1$} odds, and that the probability that at least one of the \acp{BBH} in \ac{GWTC-2} contains a \twoG{} black hole is $>99.99\%$. 

We find that for all of our assumed cluster models, the events in \ac{GWTC-2} are better fit including the hierarchical channels than when excluding those channels (equivalent to setting $V_{\mathrm{esc}}= 0~\mathrm{km\,s^{-1}}$), with the highest Bayes factors corresponding to models where the central escape velocities are $\sim 300~\mathrm{km\,s^{-1}}$. 
In Fig.~\ref{fig:vesc}, we show Bayes factors in favor of our hierarchical model versus a model with only \firstgen{} \acp{BBH}; taking the ratio of these Bayes factors gives cluster-wise Bayes factors comparing how well the data are supported by different cluster models. 
For clusters with escape velocities of $\sim 300~\mathrm{km\,s^{-1}}$, which have the highest Bayes factors, we find that $99\%$ of \firstgen{} black holes are below $\NormHeavyGaussOneGMassULNinetyNine{} \Msun{}$ ($\ZeroHeavyGaussOneGMassULNinetyNine{} \Msun{}$) and that $99\%$ of all black holes are below $\NormHeavyGaussAllMassULNinetyNine{}\Msun{}$ ($\ZeroHeavyGaussAllMassULNinetyNine\Msun{}$) using \smallspin{} (\zerospin{}). 
We infer median relative rates for \smallspin{} (\zerospin{}) of \halfgen{} and \secondgen{} versus \firstgen{} mergers of \NormHeavyGaussBranchingHalfMed{} (\ZeroHeavyGaussBranchingHalfMed{}) and \NormHeavyGaussBranchingTwoMed{} (\ZeroHeavyGaussBranchingTwoMed{}) respectively, with $99\%$ upper limits of \NormHeavyGaussBranchingHalfULNinetyNine{}{} (\ZeroHeavyGaussBranchingHalfULNinetyNine{}{}) and \NormHeavyGaussBranchingTwoULNinetyNine{}{} (\ZeroHeavyGaussBranchingTwoULNinetyNine{}{}). When accounting for selection effects, we infer median relative rates for the detected population with
\smallspin{} (\zerospin{}) of \halfgen{} and \secondgen{} versus \firstgen{} mergers of \NormHeavyGaussBranchingHalfDetMed{} (\ZeroHeavyGaussBranchingHalfDetMed{}) and \NormHeavyGaussBranchingTwoDetMed{} (\ZeroHeavyGaussBranchingTwoDetMed{}) respectively, with $99\%$ upper limits of \NormHeavyGaussBranchingHalfDetULNinetyNine{}{} (\ZeroHeavyGaussBranchingHalfDetULNinetyNine{}{}) and \NormHeavyGaussBranchingTwoDetULNinetyNine{}{} (\ZeroHeavyGaussBranchingTwoDetULNinetyNine{}{}).
When $V_{\mathrm{esc}}\sim 300~\mathrm{km\,s^{-1}}$, we find that GW190521 is most likely of \secondgen{} origin, with 1200:1 and 700:1 odds in favor of being \secondgen{} versus \firstgen{} using \smallspin{} and \zerospin{}, respectively, with both spin models favoring \secondgen{} over \halfgen{} origin at \ck{$\sim3.5{:}1$}. 
For both models, we find that GW190602, GW190620\_030421 (GW190620), GW190706, and GW190519 are most likely of \halfgen{} origin, favored over \firstgen{} origin at $>10{:}1$ odds. Using \zerospin{}, we also find that GW190517\_055101 (GW190517) is favored to be of 1G+2G over 1G+1G  origin at $>10{:}1$ odds.

\section{Conclusions}\label{sec:conclusion}

\ac{GW} observations have demonstrated that black holes merge to form more massive black holes \citep{Abbott:2016blz}. 
If these merger products form new binaries, they may again merge as a detectable \ac{GW} source. 
It is necessary to consider this hierarchical merger channel when using catalogs of \ac{GW} sources to make inferences about the physics of black hole formation. 
For example, inference of the location of the lower edge of the pair-instability mass gap, which could potentially constrain nuclear reaction rates \citep{Farmer:2020xne} or beyond Standard Model physics \citep{Croon:2020oga,Straight:2020zke,Baxter:2021swn}, using detections of black holes in the $\gtrsim 50 \Msun$ regime would be contaminated by the presence of \twoG{} black holes. 
In order to distinguish between \oneG{} and \twoG{} black holes, we must account simultaneously for the shapes of \oneG{} and \twoG{} populations and the relative rate of hierarchical mergers. 
Here, we apply the analysis of \citet{Kimball:2020opk} to $44$ \acp{BBH} in \ac{GWTC-2}, and self-consistently infer a black hole population that accounts for \firstgen{}, \halfgen{}, and \secondgen{} binary mergers, as well as the relative branching ratios between them, in order to identify candidate hierarchical mergers in the current catalog of \ac{GW} sources.

We find the following, assuming our nominal globular cluster model with $M_\mathrm{c} = 5\times10^5 \Msun$ and $r_\mathrm{c} = 1~\mathrm{pc}$: 
\begin{enumerate}
    \item The $44$ events in \ac{GWTC-2} are best modelled when allowing for hierarchical formation channels. 
    For \smallspin{} and \zerospin{}, we find Bayes factors of \ck{5} and \ck{7}, respectively, in favor of including hierarchical components. 
    \item At least one \ac{BBH} in \ac{GWTC-2} contains a \twoG{} black hole with \ck{$99\%$} and \ck{$96\%$} probability using \smallspin{} and \zerospin{}, respectively.
    \item The two binaries which are most likely to contain a \twoG{} black hole are GW190519 and GW190521, with \NormNominalGaussGWNineteenORHalf:1 and \NormNominalGaussGWTwentyOneORHalf:1 odds respectively of being \halfgen{} versus \firstgen{} assuming \smallspin{}.
    Using \zerospin{}, we find that both events have approximately equal odds of being \halfgen{} and \firstgen.
    \item The relative rates of hierarchical mergers are dependent on how the \firstgen{} spin is modelled. 
    Using \smallspin{}, the median relative merger rates of \halfgen{} and \secondgen{} to \firstgen{} mergers are inferred to be \NormNominalGaussBranchingHalfMed{} and \NormNominalGaussBranchingTwoMed{}, respectively, with $99\%$ upper limits of \NormNominalGaussBranchingHalfULNinetyNine{} and \NormNominalGaussBranchingTwoULNinetyNine{}. While using \zerospin{}, the relative rates drop to \ZeroNominalGaussBranchingHalfMed{} and \ZeroNominalGaussBranchingTwoMed{}, with $99\%$ upper limits of \ZeroNominalGaussBranchingHalfULNinetyNine{} and \ZeroNominalGaussBranchingTwoULNinetyNine{}, respectively.
    \item Using \smallspin{} (\zerospin{}), we find that $99\%$ of \firstgen{} black holes are below $\NormHeavyGaussOneGMassULNinetyNine{}\Msun{}$ ($\ZeroHeavyGaussOneGMassULNinetyNine{}\Msun{}$)  and that $99\%$ of all black holes are below $\NormHeavyGaussAllMassULNinetyNine{}\Msun{}$ ($\ZeroHeavyGaussAllMassULNinetyNine{}\Msun{}$).
\end{enumerate}
Since we do not believe that all \acp{BBH} come from a single type of cluster, we also consider a range of other typical cluster sizes, demonstrating that results depend upon the assumed escape velocity.
For a cluster model with $M_\mathrm{c}=10^6\Msun$ and $r_\mathrm{c}=0.1~\mathrm{pc}$, which has the highest Bayes factor:
\begin{enumerate}
   \item We overwhelmingly favor models including hierarchical formation channels.
   For \smallspin{} and \zerospin{}, we find Bayes factors of $1400{:}1$ and $25000{:}1$, respectively, in favor of including hierarchical components. 
   \item At least one \ac{BBH} in \ac{GWTC-2} contains a \twoG{} black hole with probability $>99.99\%$ for both \smallspin{} and \zerospin{}.
   \item GW190521 is most likely of \secondgen{} origin, with $1200{:}1$ and $700{:}1$ odds in favor of being \secondgen{} versus \firstgen{} using \smallspin{} and \zerospin{}, with both models favoring \secondgen{} over \halfgen{} origin at $\sim3.5{:}1$.
   \item We find that GW190519, GW190602, GW190620, and GW190706 are most likely of \halfgen{} origin for both \smallspin{} and \zerospin{}, favored over \firstgen{} origin at $>10{:}1$ odds, while GW190517 is favored to be of 1G+2G origin above this threshold for \zerospin{}.
   \item Using \smallspin{}, the median relative merger rates of \halfgen{} and \secondgen{} to \firstgen{} mergers are inferred to be \NormHeavyGaussBranchingHalfMed{} and \NormHeavyGaussBranchingTwoMed{}, respectively, with $99\%$ upper limits of \NormHeavyGaussBranchingHalfULNinetyNine{} and \NormHeavyGaussBranchingTwoULNinetyNine{}. 
   While using \zerospin{}, the relative rates drop slightly to \ZeroHeavyGaussBranchingHalfMed{} and \ZeroHeavyGaussBranchingTwoMed{}, with $99\%$ upper limits of \ZeroHeavyGaussBranchingHalfULNinetyNine{} and \ZeroHeavyGaussBranchingTwoULNinetyNine{}, respectively.
\end{enumerate}
Our analysis indicates that there are plausible hierarchical merger candidates in \ac{GWTC-2}, meriting further study.

There are a number of possible extensions to this analysis.
Most importantly, we have assumed that all merging binaries are formed dynamically in clusters with a specific mass and density. 
While illustrative, this is unrealistic as (i) the observed \ac{BBH} population may come from a mixture of formation channels including isolated field evolution, and (ii) dynamically formed binaries may occur in a wide range cluster types ranging from young open clusters to nuclear star clusters. 
An excess of events with aligned spin in \ac{GWTC-2} suggests that at least some binaries are assembled in the field \citep{Abbott:2020gyp}, and comparisons of observations with model predictions indicate that a mix of formation channels is probable \citep{Wong:2020ise,Zevin:2020gbd,Bouffanais:2021wcr}.
The potential for multiple formation channels could be accounted for by including an additional mixture model for dynamically formed binaries versus those formed in isolation \citep{Kimball:2020opk}. 
Previous analyses have suggested using the distribution of spin orientations or eccentricities to measure the fraction of binaries formed dynamically \citep{Vitale:2015tea,Nishizawa:2016jji,Breivik:2016ddj,Stevenson:2017dlk,Talbot:2017yur,Gondan:2018khr,Lower:2018seu,Romero-Shaw:2019itr,Abbott:2020gyp,Zevin:2020gbd}. 
Relaxing the assumption that all dynamically formed binaries form in identical environments requires a model for the distribution of globular cluster properties and other dense environments, e.g., \acp{AGN}.
It is possible that future \ac{GW} observations will allow us to directly probe the distribution of cluster masses if we obtain sufficient observations to reconstruct the population of host environments.
We leave incorporating these extensions to future work.

\acknowledgements{}
The authors thank Kyle Kremer, Carl Rodriguez, Mario Spera, and Zoheyr Doctor for their expert advice in constructing this study, and Isobel Romero-Shaw for comments on a draft manuscript. 
This research has made use of data obtained from the Gravitational Wave Open Science Center (\href{https://www.gw-openscience.org}{www.gw-openscience.org}), a service of LIGO Laboratory, the LIGO Scientific Collaboration and the Virgo Collaboration. LIGO is funded by the US National Science Foundation (NSF). Virgo is funded by the French Centre National de Recherche Scientifique (CNRS), the Italian Istituto Nazionale della Fisica Nucleare (INFN) and the Dutch Nikhef, with contributions by Polish and Hungarian institutes. 
This work is supported by the NSF Grant PHY-1607709 and through the Australian Research Council (ARC) Centre of Excellence CE170100004. 
CK is supported supported by the National Science Foundation under grant DGE-1450006.
CPLB is supported by the CIERA Board of Visitors Research Professorship. 
MZ is supported by NASA through the NASA Hubble Fellowship grant HST-HF2-51474.001-A awarded by the Space Telescope Science Institute, which is operated by the Association of Universities for Research in Astronomy, Inc., for NASA, under contract NAS5-26555. 
ET is supported through ARC Future Fellowship FT150100281 and CE170100004. 
TD acknowledges support from the Mar{\' i}a de Maeztu Unit of Excellence MDM-2016-0692, by Xunta de Galicia under project ED431C 2017/07, by Conseller{\' i}a de Educac{\' i}on, Universidade e Formac{\' i}on Profesional as Centros de Investigac{\' i}on do Sistema universitario de Galicia (ED431G 2019/05), and by FEDER. 
This research was supported in part through the computational resources from the Grail computing cluster at Northwestern University---funded through NSF PHY-1726951---and staff contributions provided for the Quest high performance computing facility at Northwestern University, which is jointly supported by the Office of the Provost, the Office for Research, and Northwestern University Information Technology. 
The authors are grateful for computational resources provided by the LIGO Laboratory and supported by NSF Grants PHY-0757058 and PHY-0823459.
This document has been assigned LIGO document number \href{https://dcc.ligo.org/LIGO-P2000466/public}{LIGO-P2000466}.

\bibliography{nextgen}

\appendix
\section{Inferred Population Hyperparameters}\label{sec:hyperparameters}

Here, we include the full sets of inferred population hyperparameter posteriors for \smallspin{} and \zerospin{}. The \firstgen{} primary mass distribution consists of two components. 
The first is a truncated power law with minimum mass $m_\mathrm{min}$, maximum mass $m_\mathrm{max}$, and power-law index $\alpha$. 
The second is a Gaussian component with mean $\mu_{m}$ and standard deviation $\sigma_{m}$. The mixing parameter $\lambda_m$ gives the fraction of BHs drawn from the Gaussian component. The mass ratio distribution is governed by a power-law with index $\beta_{q}$. 
The \firstgen{} spin distributions for \smallspin{} are modeled as truncated Gaussians, with standard deviation $\sigma_\chi \in [0.1,10]$ and mean $\mu_\chi \in [-3 \sigma_\chi,1+3\sigma_\chi]$. For \zerospin{}, we use a Beta distribution with shape parameters $\alpha_\chi > 1$ and $\beta_\chi > 1$ and allow a fraction $\lambda_0$ of the population to have spins drawn from a delta function at zero: 
\begin{align}
    \pi(\chi| \lambda_0, \alpha_\chi, \beta_\chi, \firstgen) = \lambda_0 \delta(\chi) + (1-\lambda_0) \Beta(\chi | \alpha_\chi, \beta_\chi).
    \label{eq:chi-model}
\end{align}
This zero-spin subpopulation is inspired by simulations of massive stars with efficient angular momentum transfer, where black holes in effective isolation would be born with spins of $\sim 0.01$ \citep{Qin:2018nuz,Fuller:2019sxi}, and may also describe primordial black holes \citep{DeLuca:2020qqa}.
The \halfgen{} and \secondgen{} mass and spin distributions are obtained using the transfer functions defined in \citet{Kimball:2020opk}.

\begin{figure*}
\includegraphics[width=\textwidth]{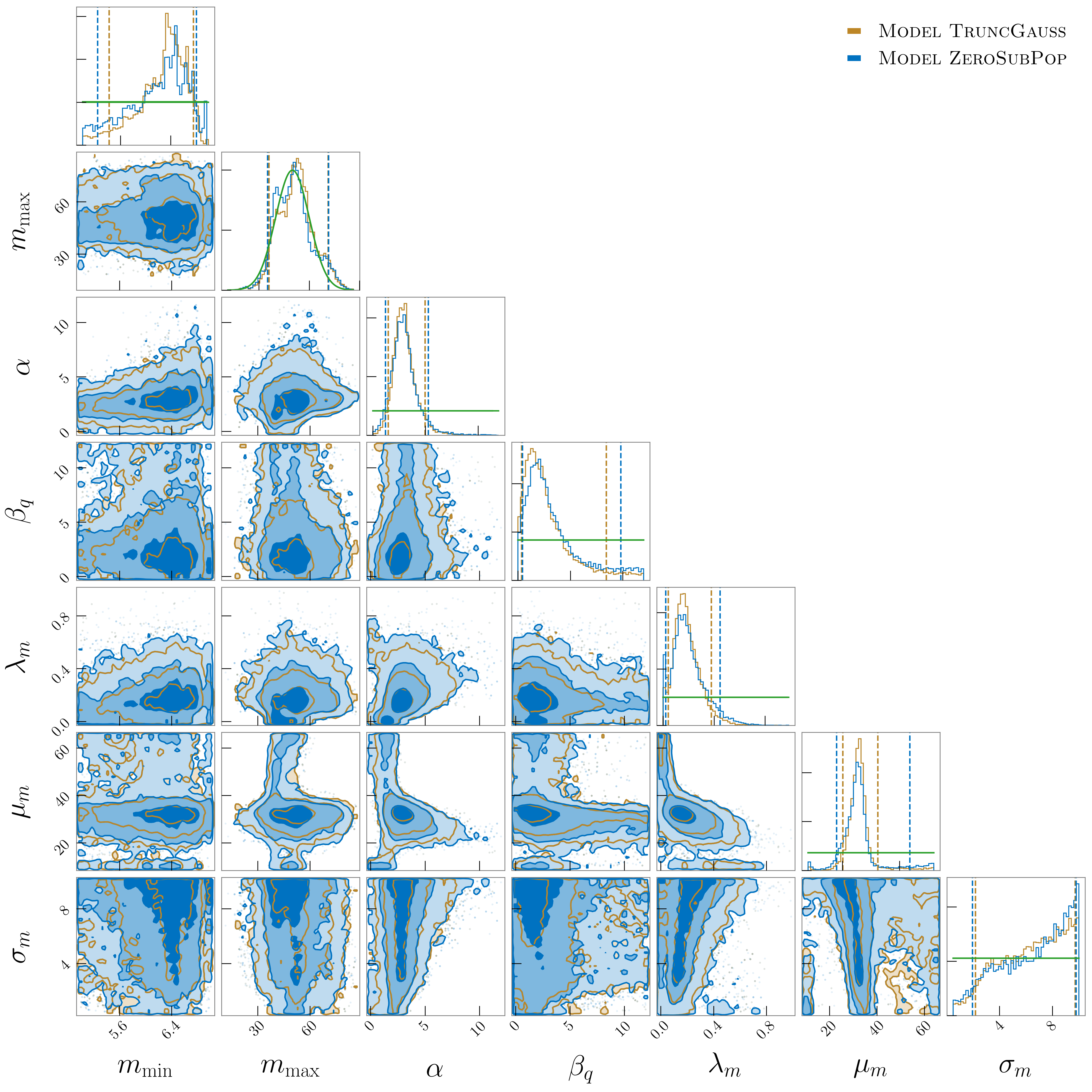}
\caption{Posterior distributions of the population hyperparameters governing the mass and mass ratio distributions. 
The dashed lines give the $90\%$ credible intervals, and the green lines indicate the priors. 
Results using \smallspin{} are plotted in orange, and results using \zerospin{} are plotted in blue.}
\label{fig:gauss_mass}
\end{figure*}

In Fig.~\ref{fig:gauss_mass}, we plot the parameters governing the mass and mass ratio distributions. 
When using the astrophysically-motivated prior on $m_\mathrm{max}$ (a Gaussian centered at $50\Msun{}$ with standard deviation $10\Msun{}$), we mostly recover this prior; we do not yet have an informative enough catalog to measure this within our phenomenological model. 
As in \citet{Kimball:2020opk}, we find that $m_\mathrm{max}$ is restricted at small values of the power-law index $\alpha$ where the mass distribution is flatter and more sensitive to the upper mass cut-off. 
We are able to place stronger constraints on the minimum black hole mass, finding $m_\mathrm{min}<\NormNominalGaussmminULNinetyNine{}$
\Msun{} and $m_\mathrm{min}<\ZeroNominalGaussmminULNinetyNine{}\Msun{}$  at the $99\%$ credible level using \smallspin{} and \zerospin{}, respectively. 
For \smallspin{}, we find that the Gaussian component of the mass spectrum is well constrained to $\mu_{m} = \NormNominalGaussmppMed^{+\NormNominalGaussmppPlus}_{-\NormNominalGaussmppMinus}\Msun{}$.
With \zerospin{}, where we infer support for lower relative hierarchical merger rates (Fig.~\ref{fig:RelativeRates}), we find a long tail at high masses when $m_\mathrm{max}$ is low, finding $\mu_{m} = \ZeroNominalGaussmppMed^{+\ZeroNominalGaussmppPlus}_{-\ZeroNominalGaussmppMinus}\Msun{}$. 
With both \smallspin{} and \zerospin{}, we find a bimodality in the relative hierarchical merger rates as seen in Fig. \ref{fig:RelativeRates}. 
The peak at higher relative rates is associated with small values of $\sigma_m$. 
When we restrict the Gaussian component to peak sharply, it is unable to accommodate the highest mass events as \firstgen{} in its tail, and they are therefore fit as hierarchical mergers. 
If we omit the five highest-mass events in \ac{GWTC-2}, this peak at higher relative hierarchical rates disappears. 
Overall, the inferred mass distributions are largely consistent between our two spin models.

\begin{figure*}
\includegraphics[width=0.45\textwidth]{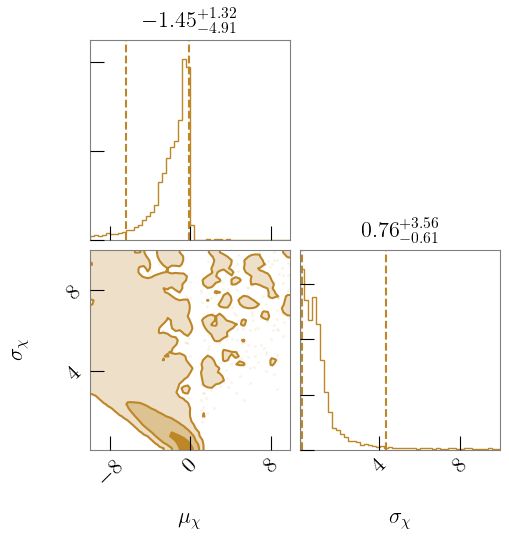}
\caption{Posterior distributions of the population hyperparameters governing the component spin distributions for \smallspin{}. 
The dashed lines give the $90\%$ credible intervals.}
\label{fig:gauss_spin_G}
\end{figure*}

\begin{figure*}
\includegraphics[width=0.45\textwidth]{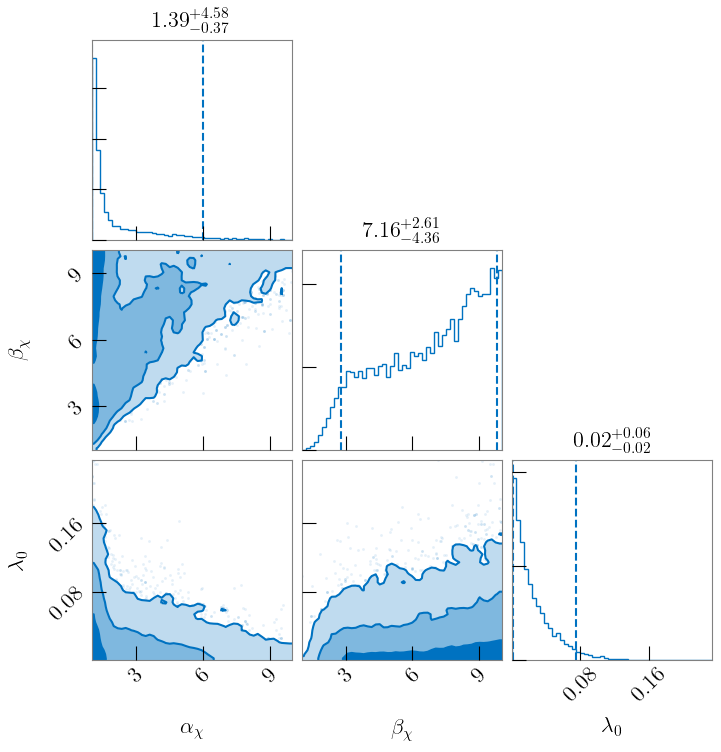}
\caption{Posterior distributions of the population hyperparameters governing the component spin distributions for \zerospin{}. 
The dashed lines give the $90\%$ credible intervals.}
\label{fig:gauss_spin_delta}
\end{figure*}

In Fig.~\ref{fig:gauss_spin_G} and Fig. ~\ref{fig:gauss_spin_delta}, we plot the parameters governing the component spin distributions for \smallspin{} and \zerospin{} respectively. 
In \smallspin{}, we find $\mu_\chi<0$, and therefore preference for spin distributions that peak at 0. In \zerospin{}, we prefer low values of $\alpha_{\chi}$, which increases support at low component spin, but find that $\beta_\chi$ is unconstrained. 
With \zerospin{}, we also find that the fraction $\lambda_0$ of black holes originating from the zero-spin subpopulation (plotted in Fig.~\ref{fig:RelativeRates}) is constrained to be less than \ZeroNominalGaussdeltachiULNinety{} (\ZeroNominalGaussdeltachiULNinetyNine{}) at the $90\%$ ($99\%$) credible level.

\begin{figure*}
\includegraphics[width=\textwidth]{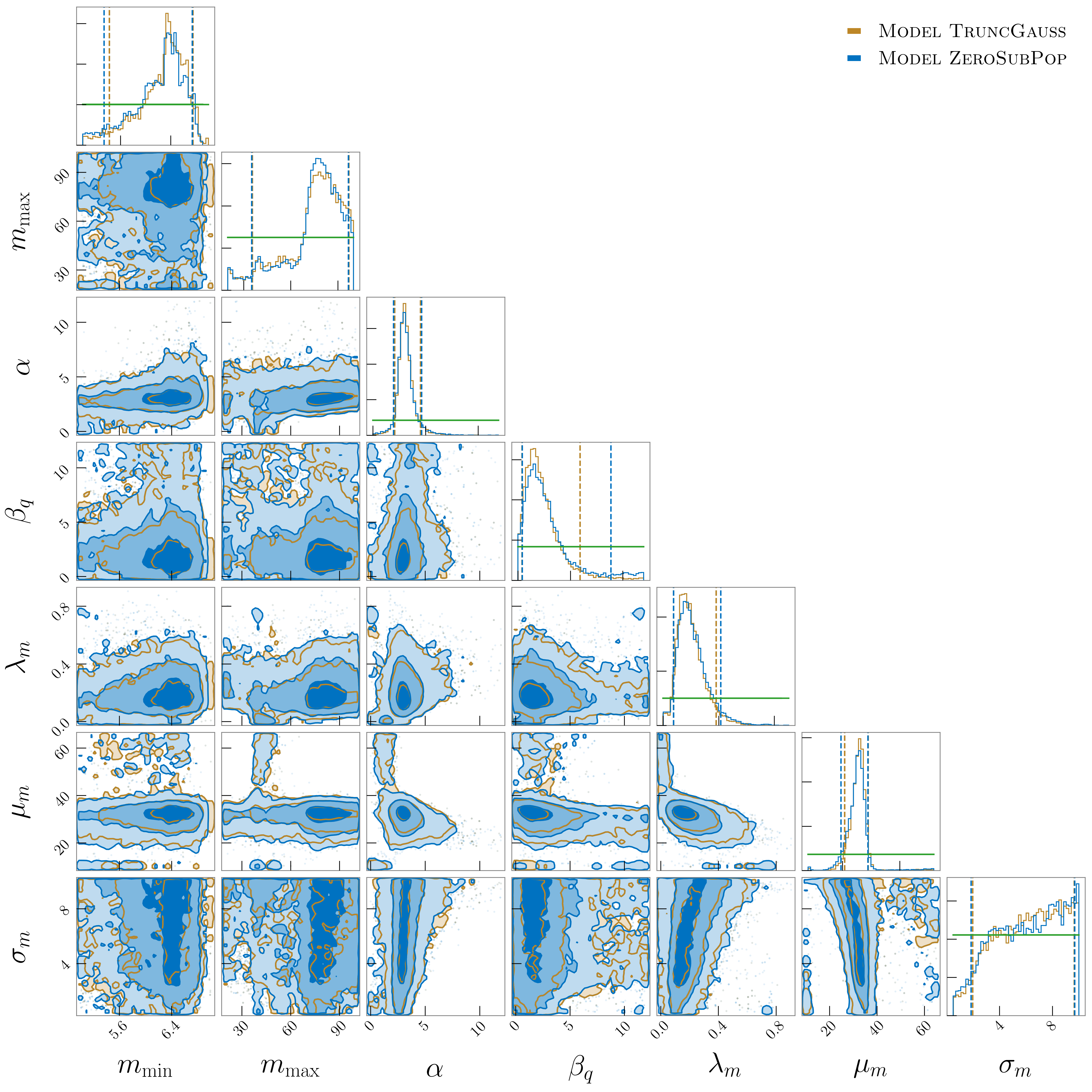}
\caption{Same as in Fig.~\ref{fig:gauss_mass} except with a flat prior on $m_\mathrm{max}$. 
The dashed lines give the $90\%$ credible intervals, and the green lines indicate the priors. 
Results using \smallspin{} are plotted in orange, and results using \zerospin{} are plotted in blue.}
\label{fig:flat_200_mass}
\end{figure*}

We plot the posteriors for the population hyperparameters governing the mass distributions inferred when we assume a flat prior on $m_\mathrm{max}$ in Fig.~\ref{fig:flat_200_mass}. 
Using the flat prior on $m_\mathrm{max}$, we no longer constrain the maximum mass cut-off, and the posterior peaks at around $\sim 80\Msun$. 
For \smallspin{} and \zerospin{}, we constrain the mean of the Gaussian component to be $\mu_{m} = \NormNominalFlatmppMed^{+\NormNominalFlatmppPlus}_{-\NormNominalFlatmppMinus}\Msun{}$ and $\mu_{m} = \ZeroNominalFlatmppMed^{+\ZeroNominalFlatmppPlus}_{-\ZeroNominalFlatmppMinus}\Msun{}$, respectively.
The preference for high $m_\mathrm{max}$ means that the high-mass tail on $\mu_m$ is no longer required to fit the more massive events in \ac{GWTC-2}, even though the relative \halfgen{} and \secondgen{} versus \firstgen{} merger rates (as well as the events-wise odds of hierarchical merger) drop by an order of magnitude under the flat prior on $m_\mathrm{max}$. 

\begin{figure*}
\includegraphics[width=0.45\textwidth]{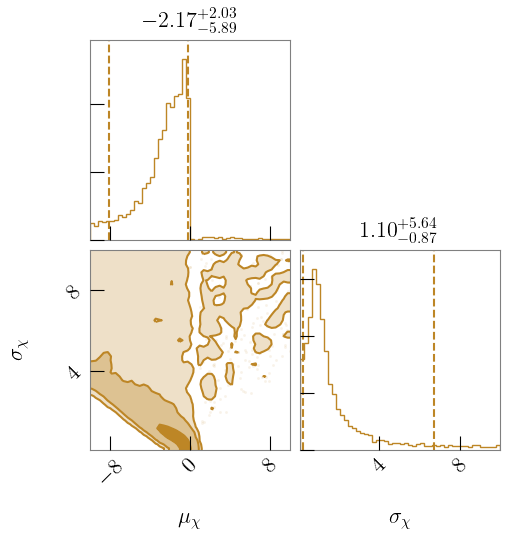}
\caption{Same as in Fig.~\ref{fig:gauss_spin_G}, except with a flat prior on $m_\mathrm{max}$. 
The dashed lines give the $90\%$ credible intervals. }
\label{fig:flat_200_spin_G}
\end{figure*}

\begin{figure*}
\includegraphics[width=0.45\textwidth]{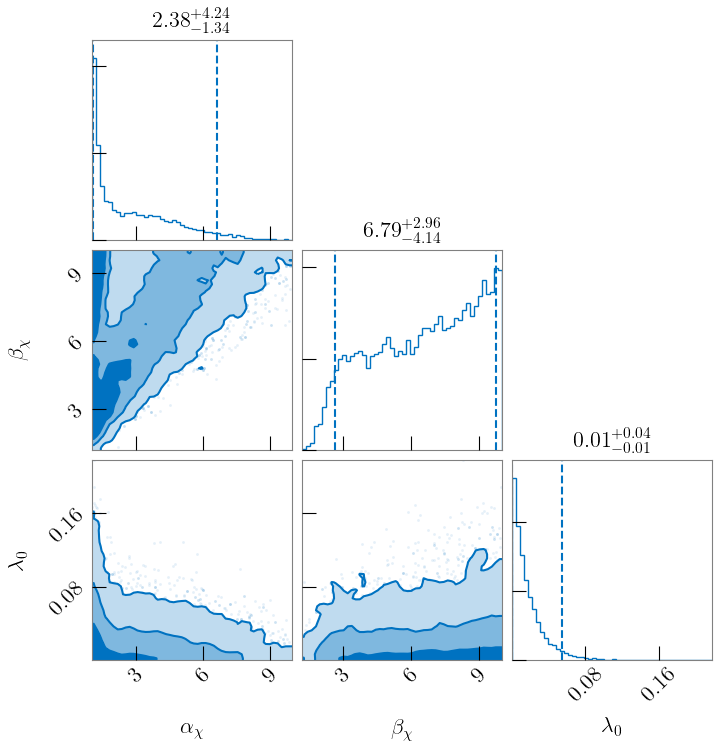}
\caption{Same as in Fig.~\ref{fig:gauss_spin_delta} except with a flat prior on $m_\mathrm{max}$. 
The dashed lines give the $90\%$ credible intervals. }
\label{fig:flat_200_spin_delta}
\end{figure*}

In Fig.~\ref{fig:flat_200_spin_G} and Fig.~\ref{fig:flat_200_spin_delta}, we plot the posteriors of the population hyperparameters governing the component spin distributions inferred when we assume a flat prior on $m_\mathrm{max}$. 
We find that the inferred spin distributions are consistent across choices of prior on $m_\mathrm{max}$. 
For \zerospin{} the fraction $\lambda_0$ of \acp{BBH} originating from the zero-spin subpopulation is constrained to be less than \ZeroNominalFlatdeltachiULNinety{} (\ZeroNominalFlatdeltachiULNinetyNine{}) at the $90\%$ ($99\%$) credible level, and is still consistent with $\lambda_0=0$. 
We also find that $\alpha_\chi$ and $\beta_\chi$ are less constrained; when we allow for high values of $m_\mathrm{max}$, it is easier to explain higher mass systems as \firstgen{}, and hence we do not require a low-spin population to enable high relative hierarchical merger rates.


\end{document}

%% file: updated_final_pe_macros.tex
\def\NormNominalGaussPAtLeastOneTwoG{\ensuremath{99}}
\def\NormNominalGaussGWNineteenORHalf{\ensuremath{1.2}}

\def\NormNominalGaussGWTwentyOneORHalf{\ensuremath{2.0}}
\def\NormNominalGaussGWTwentyOneORTwo{\ensuremath{1.2}}

\def\NormNominalGaussmminULNinetyNine{\ensuremath{6.8}}

\def\NormNominalGaussmppMed{\ensuremath{32.0}}

\def\NormNominalGaussmppMinus{\ensuremath{6.8}}
\def\NormNominalGaussmppPlus{\ensuremath{8.5}}

\def\NormNominalGaussBranchingHalfMed{\ensuremath{5.8\times 10^{-3}}}

\def\NormNominalGaussBranchingHalfULNinetyNine{\ensuremath{0.12}}

\def\NormNominalGaussBranchingTwoMed{\ensuremath{1.7\times 10^{-5}}}

\def\NormNominalGaussBranchingTwoULNinetyNine{\ensuremath{7.6\times 10^{-3}}}

\def\NormNominalGaussBranchingHalfDetMed{\ensuremath{0.01}}

\def\NormNominalGaussBranchingHalfDetULNinetyNine{\ensuremath{0.23}}

\def\NormNominalGaussBranchingTwoDetMed{\ensuremath{7.0\times 10^{-5}}}

\def\NormNominalGaussBranchingTwoDetULNinetyNine{\ensuremath{0.03}}

\def\NormNominalGaussvt1Low{\ensuremath{0.3}}
\def\NormNominalGaussvt1Med{\ensuremath{0.4}}
\def\NormNominalGaussvt1High{\ensuremath{0.6}}
\def\NormNominalGaussvt1Minus{\ensuremath{0.1}}
\def\NormNominalGaussvt1Plus{\ensuremath{0.2}}
\def\NormNominalGaussvt1ULNinetyNine{\ensuremath{0.7}}
\def\NormNominalGaussvt1ULNinety{\ensuremath{0.5}}
\def\NormNominalGaussvt15Low{\ensuremath{0.5}}
\def\NormNominalGaussvt15Med{\ensuremath{0.8}}
\def\NormNominalGaussvt15High{\ensuremath{1.2}}
\def\NormNominalGaussvt15Minus{\ensuremath{0.3}}
\def\NormNominalGaussvt15Plus{\ensuremath{0.4}}
\def\NormNominalGaussvt15ULNinetyNine{\ensuremath{1.4}}
\def\NormNominalGaussvt15ULNinety{\ensuremath{1.1}}
\def\NormNominalGaussvt2Low{\ensuremath{1.1}}
\def\NormNominalGaussvt2Med{\ensuremath{1.6}}
\def\NormNominalGaussvt2High{\ensuremath{2.3}}
\def\NormNominalGaussvt2Minus{\ensuremath{0.5}}
\def\NormNominalGaussvt2Plus{\ensuremath{0.7}}
\def\NormNominalGaussvt2ULNinetyNine{\ensuremath{2.7}}
\def\NormNominalGaussvt2ULNinety{\ensuremath{2.1}}
\def\NormNominalGaussOneGSpinULNinety{\ensuremath{0.65}}
\def\NormNominalGaussOneGMassULNinetyNine{\ensuremath{47}}
\def\NormNominalGaussAllMassULNinetyNine{\ensuremath{49}}

\def\NormNominalFlatmppMed{\ensuremath{32.1}}

\def\NormNominalFlatmppMinus{\ensuremath{6.1}}
\def\NormNominalFlatmppPlus{\ensuremath{4.2}}

\def\NormNominalFlatOneGMassULNinetyNine{\ensuremath{49}}
\def\NormNominalFlatAllMassULNinetyNine{\ensuremath{51}}
\def\ZeroNominalGaussPAtLeastOneTwoG{\ensuremath{96}}
\def\ZeroNominalGaussGWNineteenORHalf{\ensuremath{1.1}}

\def\ZeroNominalGaussGWTwentyOneORHalf{\ensuremath{1.0}}

\def\ZeroNominalGaussmminULNinetyNine{\ensuremath{7.0}}

\def\ZeroNominalGaussmppMed{\ensuremath{31.5}}

\def\ZeroNominalGaussmppMinus{\ensuremath{9.0}}
\def\ZeroNominalGaussmppPlus{\ensuremath{23.0}}

\def\ZeroNominalGaussdeltachiULNinetyNine{\ensuremath{0.12}}
\def\ZeroNominalGaussdeltachiULNinety{\ensuremath{0.06}}

\def\ZeroNominalGaussBranchingHalfMed{\ensuremath{5.3\times 10^{-3}}}

\def\ZeroNominalGaussBranchingHalfULNinetyNine{\ensuremath{0.04}}

\def\ZeroNominalGaussBranchingTwoMed{\ensuremath{1.4\times 10^{-5}}}

\def\ZeroNominalGaussBranchingTwoULNinetyNine{\ensuremath{9.8\times 10^{-4}}}

\def\ZeroNominalGaussBranchingHalfDetMed{\ensuremath{0.01}}

\def\ZeroNominalGaussBranchingHalfDetULNinetyNine{\ensuremath{0.08}}

\def\ZeroNominalGaussBranchingTwoDetMed{\ensuremath{5.7\times 10^{-5}}}

\def\ZeroNominalGaussBranchingTwoDetULNinetyNine{\ensuremath{4.0\times 10^{-3}}}

\def\ZeroNominalGaussvt1Low{\ensuremath{0.3}}
\def\ZeroNominalGaussvt1Med{\ensuremath{0.4}}
\def\ZeroNominalGaussvt1High{\ensuremath{0.6}}
\def\ZeroNominalGaussvt1Minus{\ensuremath{0.1}}
\def\ZeroNominalGaussvt1Plus{\ensuremath{0.2}}
\def\ZeroNominalGaussvt1ULNinetyNine{\ensuremath{0.7}}
\def\ZeroNominalGaussvt1ULNinety{\ensuremath{0.6}}
\def\ZeroNominalGaussvt15Low{\ensuremath{0.5}}
\def\ZeroNominalGaussvt15Med{\ensuremath{0.8}}
\def\ZeroNominalGaussvt15High{\ensuremath{1.3}}
\def\ZeroNominalGaussvt15Minus{\ensuremath{0.3}}
\def\ZeroNominalGaussvt15Plus{\ensuremath{0.5}}
\def\ZeroNominalGaussvt15ULNinetyNine{\ensuremath{1.4}}
\def\ZeroNominalGaussvt15ULNinety{\ensuremath{1.1}}
\def\ZeroNominalGaussvt2Low{\ensuremath{1.1}}
\def\ZeroNominalGaussvt2Med{\ensuremath{1.6}}
\def\ZeroNominalGaussvt2High{\ensuremath{2.4}}
\def\ZeroNominalGaussvt2Minus{\ensuremath{0.5}}
\def\ZeroNominalGaussvt2Plus{\ensuremath{0.8}}
\def\ZeroNominalGaussvt2ULNinetyNine{\ensuremath{2.8}}
\def\ZeroNominalGaussvt2ULNinety{\ensuremath{2.2}}
\def\ZeroNominalGaussOneGSpinULNinety{\ensuremath{0.50}}
\def\ZeroNominalGaussOneGMassULNinetyNine{\ensuremath{47}}
\def\ZeroNominalGaussAllMassULNinetyNine{\ensuremath{48}}

\def\ZeroNominalFlatmppMed{\ensuremath{31.6}}

\def\ZeroNominalFlatmppMinus{\ensuremath{7.3}}
\def\ZeroNominalFlatmppPlus{\ensuremath{4.5}}

\def\ZeroNominalFlatdeltachiULNinetyNine{\ensuremath{0.09}}
\def\ZeroNominalFlatdeltachiULNinety{\ensuremath{0.04}}

\def\ZeroNominalFlatOneGMassULNinetyNine{\ensuremath{49}}
\def\ZeroNominalFlatAllMassULNinetyNine{\ensuremath{50}}

\def\NormHeavyGaussBranchingHalfMed{\ensuremath{0.15}}

\def\NormHeavyGaussBranchingHalfULNinetyNine{\ensuremath{0.29}}

\def\NormHeavyGaussBranchingTwoMed{\ensuremath{0.01}}

\def\NormHeavyGaussBranchingTwoULNinetyNine{\ensuremath{0.04}}

\def\NormHeavyGaussBranchingHalfDetMed{\ensuremath{0.3}}

\def\NormHeavyGaussBranchingHalfDetULNinetyNine{\ensuremath{0.57}}

\def\NormHeavyGaussBranchingTwoDetMed{\ensuremath{0.05}}

\def\NormHeavyGaussBranchingTwoDetULNinetyNine{\ensuremath{0.18}}

\def\NormHeavyGaussvt1Low{\ensuremath{0.2}}
\def\NormHeavyGaussvt1Med{\ensuremath{0.4}}
\def\NormHeavyGaussvt1High{\ensuremath{0.5}}
\def\NormHeavyGaussvt1Minus{\ensuremath{0.1}}
\def\NormHeavyGaussvt1Plus{\ensuremath{0.2}}
\def\NormHeavyGaussvt1ULNinetyNine{\ensuremath{0.6}}
\def\NormHeavyGaussvt1ULNinety{\ensuremath{0.5}}
\def\NormHeavyGaussvt15Low{\ensuremath{0.5}}
\def\NormHeavyGaussvt15Med{\ensuremath{0.7}}
\def\NormHeavyGaussvt15High{\ensuremath{1.1}}
\def\NormHeavyGaussvt15Minus{\ensuremath{0.2}}
\def\NormHeavyGaussvt15Plus{\ensuremath{0.4}}
\def\NormHeavyGaussvt15ULNinetyNine{\ensuremath{1.3}}
\def\NormHeavyGaussvt15ULNinety{\ensuremath{1.0}}
\def\NormHeavyGaussvt2Low{\ensuremath{1.0}}
\def\NormHeavyGaussvt2Med{\ensuremath{1.5}}
\def\NormHeavyGaussvt2High{\ensuremath{2.2}}
\def\NormHeavyGaussvt2Minus{\ensuremath{0.5}}
\def\NormHeavyGaussvt2Plus{\ensuremath{0.7}}
\def\NormHeavyGaussvt2ULNinetyNine{\ensuremath{2.6}}
\def\NormHeavyGaussvt2ULNinety{\ensuremath{2.1}}

\def\NormHeavyGaussOneGMassULNinetyNine{\ensuremath{40}}
\def\NormHeavyGaussAllMassULNinetyNine{\ensuremath{67}}

\def\ZeroHeavyGaussBranchingHalfMed{\ensuremath{0.14}}

\def\ZeroHeavyGaussBranchingHalfULNinetyNine{\ensuremath{0.25}}

\def\ZeroHeavyGaussBranchingTwoMed{\ensuremath{9.2\times 10^{-3}}}

\def\ZeroHeavyGaussBranchingTwoULNinetyNine{\ensuremath{0.03}}

\def\ZeroHeavyGaussBranchingHalfDetMed{\ensuremath{0.26}}

\def\ZeroHeavyGaussBranchingHalfDetULNinetyNine{\ensuremath{0.48}}

\def\ZeroHeavyGaussBranchingTwoDetMed{\ensuremath{0.04}}

\def\ZeroHeavyGaussBranchingTwoDetULNinetyNine{\ensuremath{0.13}}

\def\ZeroHeavyGaussvt1Low{\ensuremath{0.2}}
\def\ZeroHeavyGaussvt1Med{\ensuremath{0.4}}
\def\ZeroHeavyGaussvt1High{\ensuremath{0.6}}
\def\ZeroHeavyGaussvt1Minus{\ensuremath{0.1}}
\def\ZeroHeavyGaussvt1Plus{\ensuremath{0.2}}
\def\ZeroHeavyGaussvt1ULNinetyNine{\ensuremath{0.6}}
\def\ZeroHeavyGaussvt1ULNinety{\ensuremath{0.5}}
\def\ZeroHeavyGaussvt15Low{\ensuremath{0.5}}
\def\ZeroHeavyGaussvt15Med{\ensuremath{0.7}}
\def\ZeroHeavyGaussvt15High{\ensuremath{1.1}}
\def\ZeroHeavyGaussvt15Minus{\ensuremath{0.3}}
\def\ZeroHeavyGaussvt15Plus{\ensuremath{0.4}}
\def\ZeroHeavyGaussvt15ULNinetyNine{\ensuremath{1.3}}
\def\ZeroHeavyGaussvt15ULNinety{\ensuremath{1.0}}
\def\ZeroHeavyGaussvt2Low{\ensuremath{1.0}}
\def\ZeroHeavyGaussvt2Med{\ensuremath{1.6}}
\def\ZeroHeavyGaussvt2High{\ensuremath{2.3}}
\def\ZeroHeavyGaussvt2Minus{\ensuremath{0.5}}
\def\ZeroHeavyGaussvt2Plus{\ensuremath{0.7}}
\def\ZeroHeavyGaussvt2ULNinetyNine{\ensuremath{2.6}}
\def\ZeroHeavyGaussvt2ULNinety{\ensuremath{2.1}}

\def\ZeroHeavyGaussOneGMassULNinetyNine{\ensuremath{40}}
\def\ZeroHeavyGaussAllMassULNinetyNine{\ensuremath{66}}